\documentclass[aps,prl,reprint,superscriptaddress]{revtex4-2}

\usepackage{ulem}
\usepackage{xcolor}
\usepackage[colorlinks=true]{hyperref}
\usepackage{booktabs}
\usepackage{soul}

\usepackage{amsmath,amssymb,color,epsfig}
\allowdisplaybreaks[4]

\newcommand{\be}{\begin{equation}}
\newcommand{\ee}{\end{equation}}
\newcommand{\bea}{\setlength\arraycolsep{2pt} \begin{eqnarray}}
\newcommand{\eea}{\end{eqnarray}}
\newcommand{\nn}{\nonumber}

\begin{document}

\hypersetup{
    linkcolor=blue,
    citecolor=red,
    urlcolor=magenta
}


\title{Neutron stars more compact than black holes as a probe of strong-field gravity}


\author{Shoulong Li}
\email[]{shoulongli@hunnu.edu.cn}
\affiliation{Department of Physics, Synergetic Innovation Center for Quantum Effect and Applications, and Institute of Interdisciplinary Studies, Hunan Normal University, Changsha, 410081, China}

\author{H. L\"u}
\email[]{mrhonglu@gmail.com}
\affiliation{Center for Joint Quantum Studies and Department of Physics, School of Science, Tianjin University, 135 Yaguan Road, Tianjin 300350, China}
\affiliation{Joint School of National University of Singapore and Tianjin University, International Campus of Tianjin University, Binhai New City, Fuzhou 350207, China}

\author{Yong Gao}
\email[]{gaoyong.physics@pku.edu.cn}
\affiliation{Max Planck Institute for Gravitational Physics (Albert Einstein Institute), Potsdam, 14476, Germany}
\affiliation{Kavli Institute for Astronomy and Astrophysics, Peking University, Beijing, 100871, China}
\affiliation{Department of Astronomy, School of Physics, Peking University, Beijing, 100871, China}

\author{Rui Xu}
\email[]{xuru@tsinghua.edu.cn}
\affiliation{Department of Astronomy, Tsinghua University, Beijing, 100084, China}
\affiliation{Kavli Institute for Astronomy and Astrophysics, Peking University, Beijing, 100871, China}

\author{Lijing Shao}
\email[]{lshao@pku.edu.cn}
\affiliation{Kavli Institute for Astronomy and Astrophysics, Peking University, Beijing, 100871, China}
\affiliation{National Astronomical Observatories, Chinese Academy of Sciences, Beijing, 100012, China}

\author{Hongwei Yu}
\email[Corresponding author: ]{hwyu@hunnu.edu.cn}
\affiliation{Department of Physics, Synergetic Innovation Center for Quantum Effect and Applications, and Institute of Interdisciplinary Studies, Hunan Normal University, Changsha, 410081, China}


\date{\today}

\begin{abstract}
Probing gravity in its strongest regime is a central goal of modern physics, as the nature of the most compact objects reflects fundamental aspects of Einstein's theory of general relativity (GR). In GR, black holes are regarded as the most compact objects in the Universe. Here, for the first time, we demonstrate that stable stellar configurations more compact than black holes can arise when neutron-star equations of state are embedded in quasi-topological gravity, a class of higher-curvature extensions of GR. We construct such ultra-compact stars, analyze their macroscopic properties, and establish their stability against radial perturbations, confirming their physical plausibility. We further identify potential observational signatures to distinguish these stars from black holes, most notably gravitational-wave echoes whose detectability could provide direct evidence of physics beyond Einstein's GR in the strong-field regime.
\end{abstract}

\maketitle

\textit{Introduction}
General relativity (GR) has withstood all precision tests in weak-field regimes and many in strong-field regimes, establishing it as the most robust and predictive theory of gravity to date~\cite{Will:2014kxa,Berti:2015itd,Barack:2018yly,Freire:2024adf,Psaltis:2008bb}. However, from a field-theoretic perspective, GR is neither ultraviolet-complete nor renormalizable, suggesting that higher-curvature corrections could be required at high energy scales or in extreme gravitational environments~\cite{Stelle:1976gc,Starobinsky:1980te,Donoghue:1994dn}. Observationally, gravitational waves from compact binary coalescences~\cite{LIGOScientific:2016aoc} and photon spheres around supermassive black holes~\cite{EventHorizonTelescope:2019dse} have been detected, yet some key predictions of GR in the strong-field regime, most notably, the existence of event horizons, still lack direct confirmation~\cite{Cardoso:2016rao,Cardoso:2016oxy,Cunha:2018gql,Cardoso:2017cqb,Cardoso:2019rvt,Cunha:2018acu,Herdeiro:2021lwl}. These theoretical challenges and observational uncertainties have driven increasing interest in probing gravity beyond GR in the strong-field regime~\cite{Damour:1993hw,Doneva:2017bvd,Silva:2017uqg,Antoniou:2017acq,Lu:2015cqa,Liu:2020yqa,Mendes:2018qwo}, while maintaining consistency with its well-tested predictions in the weak-field limit.

Neutron stars, due to their extreme densities and strong gravitational fields, offer an exceptional laboratory for testing strong-field gravity~\cite{Freire:2024adf,Shao:2022koz,Silva:2024cit,Olmo:2019flu,Yagi:2016bkt}.  A  typical neutron star with a mass of 1.4 solar masses has a radius of about 11 kilometers~\cite{Capano:2019eae}, corresponding to a compactness of about 0.19, where compactness is defined as the ratio of mass to radius in geometric units ($G = c = 1$). More massive neutron stars usually exhibit higher compactness, approaching but not exceeding the black hole limit of $0.5$. This limit arises because, as mass approaches the maximum supported by the equation of state (EOS), the star becomes more susceptible to collapse. Additionally, rotational effects further limit achievable compactness:  As the mass approaches its maximum value, the moment of inertia generally decreases with increasing compactness, resulting in higher angular velocity at a given angular momentum. Highly compact, rapidly rotating neutron stars typically develop an ergoregion, resulting in linear instabilities known as ergoregion instability (or w-mode instability)~\cite{Friedman:1978ygc,Comins1,Yoshida1,Kokkotas:2002sf}, triggering collapse to black holes. Although certain conditions, such as slow rotation~\cite{Cardoso:2017cqb,Cardoso:2019rvt,Friedman:1978wla,Cardoso:2007az,Brito:2015oca,LISA:2022kgy} or unusual matter properties~\cite{Raposo:2018rjn,Alho:2022bki,Maggio:2017ivp,Maggio:2018ivz}, might theoretically permit neutron stars to approach the black hole compactness threshold, standard GR forbids stable horizonless stars more compact than black holes of identical mass~\cite{Penrose:1964wq,Misner:1973prb}, where the exterior spacetime coincides with the black hole, such that any excess in compactness inevitably results in the emergence of an event horizon and a singularity.

However, this compactness restriction is deeply tied to the structure of GR, and need not apply if the theory is suitably extended. In this Letter, we explore a scenario in which neutron stars can surpass black hole compactness (i.e., have radii smaller than those of equal-mass black holes), within an extended theory of gravity known as quasi-topological gravity (QTG)~\cite{Li:2017ncu,Hennigar:2017ego}. In the QTG framework, we demonstrate the existence of horizonless, non-singular neutron stars more compact than black holes, highlighting how these objects provide distinctive observational signatures, particularly gravitational-wave echoes, as powerful probes of strong-field deviations from GR.

\textit{Stellar configuration} 
Inspired by prior studies indicating that slowly rotating ultra-compact objects can remain stable over long timescales~\cite{Cardoso:2017cqb,Cardoso:2019rvt,Friedman:1978wla,Cardoso:2007az,Brito:2015oca,LISA:2022kgy}, we assume slow rotation to explore neutron stars surpassing black hole compactness within QTG. Under this assumption, the spacetime remains approximately spherically symmetric and is described by: 
\begin{eqnarray}
ds^2&=&-h(r)dt^2+f(r)^{-1}dr^2+r^2(d\theta^2+\sin^2\theta d\phi^2) \nn \\
&\quad&- 2 \epsilon (\Omega-w(r)) r^2 \sin^2\theta dt d\phi \,, \label{ansatz}
\end{eqnarray}
in standard Schwarzschild-like coordinates $(t, r, \theta, \phi)$, where $\epsilon$ is a bookkeeping parameter characterizing slow-rotation. Here, $\Omega$ denotes the angular velocity observed inside the star, and $w$ represents  the frame-dragging angular velocity. To avoid the GR-predicted inevitable collapse into a black hole at high compactness, we consider QTG~\cite{Li:2017ncu,Hennigar:2017ego}, described by the Lagrangian $L$ given by
\begin{equation}
L= {\cal R} + \lambda ( {\cal R}^3-6{\cal R}{\cal R}_{\mu\nu}{\cal R}^{\mu\nu}+8{{\cal R}_\mu}^\nu{{\cal R}_\nu}^\gamma{{\cal R}_\gamma}^\mu ) \,,  \label{qtg}
\end{equation}
where ${\cal R}$ and ${\cal R}_{\mu\nu}$ are the Ricci scalar and Ricci tensor respectively, and $\lambda$ is the QTG coupling constant. 

QTG offers two key advantages for neutron star modeling. Firstly, it  admits a broader class of spherically symmetric vacuum solutions than GR, including the Schwarzschild spacetime as a special case~\cite{Li:2017ncu}, allowing the external spacetime of a neutron star to deviate from that of a black hole and enabling horizonless stars surpassing black hole compactness.  Secondly, QTG recovers GR in weak-field regimes~\cite{Li:2017ncu,Hennigar:2017ego}, ensuring consistency with observational constraints without the introduction of additional modes in the linear spectrum of the weak-field vacuum region, such as ghost-like massive spin-2 modes~\cite{Stelle:1976gc}. 

\begin{figure}[h]
\centering
\includegraphics[width=0.45\textwidth]{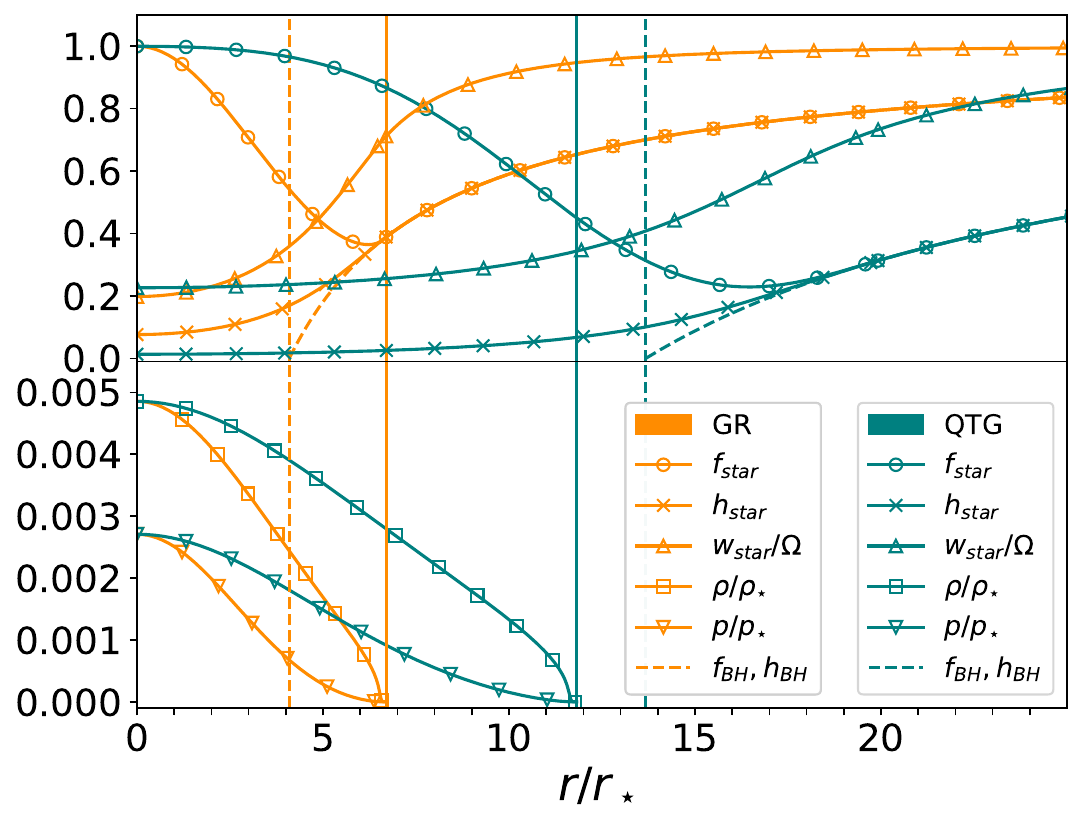}
\caption{Spacetime and matter profiles for slowly rotating stars (solid) and black holes (dashed) of the same mass in both GR (yellow) and QTG (green, $\lambda = 500 \lambda_\star$). Vertical lines denote the stellar radius (solid) and black hole horizon (dashed). The SLy EOS is used with central density $\rho_0 = 4.86 \times 10^{-3} \rho_\star$ ($3 \times 10^{15} \textup{g/cm}^3$). In GR, the star has mass $M=2.05 M_{\odot}$ and moment of inertia $I= 43.27 I_\star$, whereas in QTG, these values increase to $6.84 M_{\odot}$ and $1056.35 I_\star$. }\label{fig:solution}
\end{figure}

We now construct slowly rotating stellar solutions within this framework. Starting with the metric ansatz in  Eq.~(\ref{ansatz}) and a prescribed EOS of the form  $p(r)=E(\rho(r))$,  we solve the gravitational field equations order by order in  the slow-rotation parameter $\epsilon$. This yields the metric functions  $h$, $f$ and $w$, along with the pressure $p$ and density $\rho$, for stars characterized by a range of central densities.  Figure~\ref{fig:solution} displays the spacetime and matter profiles for a representative stellar model constructed using the realistic SLy (Skyrme Lyon) EOS~\cite{Douchin:2001sv,Haensel:2004nu}, under a typical central density, in both GR and QTG. In QTG, the interior spacetime geometry is significantly modified:  the metric functions $h$ and $f$  differ at the stellar surface due to continuity conditions, but rapidly converge to equality prior to entering the weak-field regime, where $h = f = 1 - 2M/r$ remains the universal form.
 
The presence of higher-curvature corrections in QTG causes both the pressure and density to decrease more slowly than in GR. As a result, the pressure vanishes at a larger radius, leading to a larger stellar radius, while the slower decline in density results in a higher total mass. Despite these modifications, equilibrium is still achieved through the balance between gravitational attraction and the outward pressure gradient.

For a QTG model with coupling constant $\lambda = 500 \lambda_\star$, the resulting stellar configuration reaches  a mass of $6.84 M_\odot$, a radius of $11.82 r_\star$, and a moment of inertia of $1056.35 I_\star$ (shown as solid green curves). For comparison, a QTG black hole with the same mass and angular velocity has an event horizon at $13.68\,r_\star$ (dashed green curves). Notably, the Schwarzschild solution remains the unique black hole solution in QTG at zeroth order in slow rotation, as shown in Appendix~A. Here, the symbol ``$\star$" used as a subscript represents the dimension of various physical quantities, $M_\odot  \sim r_\star \sim \rho_\star^{-1/2}  \sim p_\star^{-1/2} \sim I_\star^{1/3} \sim \lambda_\star^{1/4} $, with the solar mass $M_{\odot}$ serving as the reference unit~\cite{footnote1}. The  compactness of the star  ${\cal C}= M/R\approx 0.58$ thus exceeds  black hole  threshold ${\cal C}= 0.5$. This enhancement in compactness arises from a key structural effect: as $\lambda$ increases, the stellar mass grows faster than the radius, a trend analyzed in detail in Appendix~B.

\begin{figure}[h]
\centering
\includegraphics[width=0.95\linewidth]{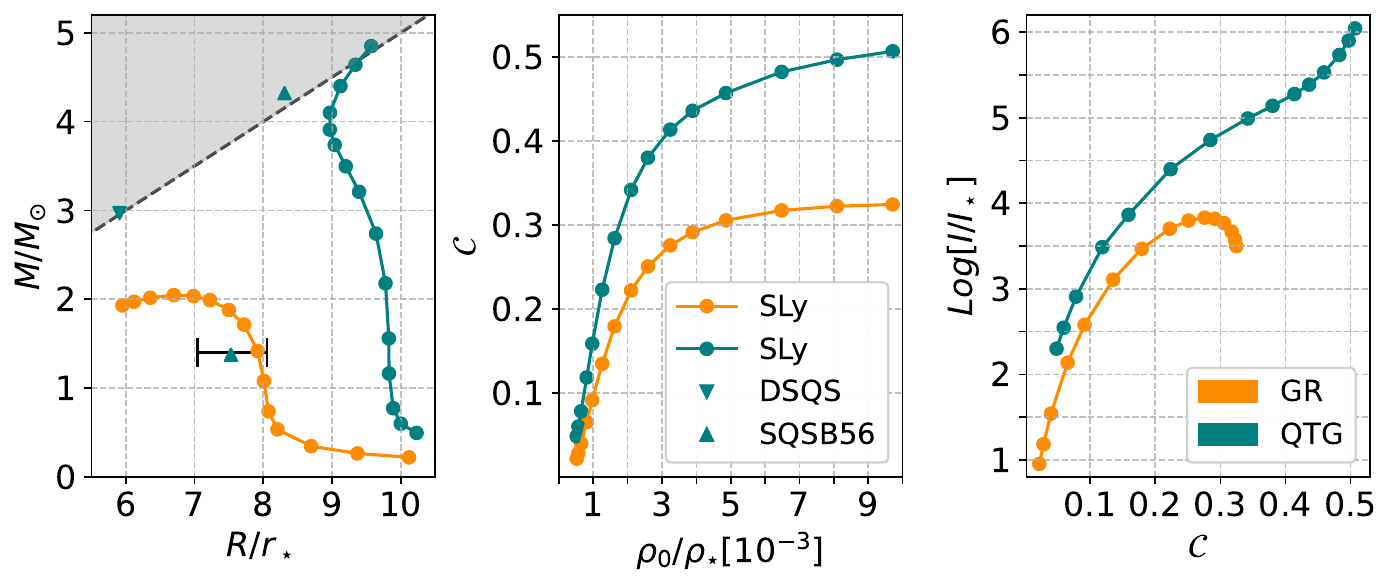}
\caption{Macroscopic properties of stellar models in GR (yellow) and QTG (green) using different EOSs. Shown are $M-R$ (left), ${\cal C}-\rho_0$ (middle), and $I-{\cal C}$ (right) relations. Coupling constants $\lambda$ for SLy, DSQS, and SQSB56 EOSs are set to $ 65 \lambda_\star$, $15 \lambda_\star$, and $100 \lambda_\star$, respectively. The error bar marks a $1.4 M_{\odot}$ neutron star with radius $R = 7.45^{+0.61}_{-0.41} r_\star$ ($11.0^{+0.9}_{-0.6} \textup{km}$, 90\% credible interval)~\cite{Capano:2019eae}. The black dashed line shows the black hole mass-radius relation and the grey region indicates compactness exceeding the black hole limit.}\label{fig:mrcrhoic}
\end{figure}

\textit{Global structure} 
Having established the existence of stellar configurations in QTG that exceed the compactness of black holes,  we now examine their global structure by analyzing macroscopic relationships among the mass $M$, radius $R$, compactness ${\cal C}$, and moment of inertia $I$. To this end, we numerically solve the stellar equilibrium equations across a range of central densities $\rho_0$ and coupling constants $\lambda$, using various realistic EOSs. Figure~\ref{fig:mrcrhoic} presents the resulting mass-radius ($M-R$), compactness-density (${\cal C}-\rho_0$), and moment of inertia-compactness ($I-{\cal C}$) relations.  These plots serve to characterize the physical viability and rotational behavior of ultra-compact stars within QTG, in contrast to those described by GR.

A key distinction from GR emerges in  the $M-R$ and ${\cal C}-\rho_0$ relations shown in the left and middle panels of Fig.~\ref{fig:mrcrhoic}: while GR imposes an upper limit on the stellar mass at a given EOS  (SLy), no such limit appears within QTG ($\lambda = 65 \lambda_\star$) over the typical range of central densities considered. At low densities, QTG stars behave similarly to their GR counterparts, with radii decreasing as mass and compactness increase. However, at higher central densities, QTG stars exhibit an unconventional trend: their radii increase with mass, reminiscent of black hole behavior. Yet unlike black holes, which exhibit $R \propto M$ and  maintain fixed compactness ${\cal C} =0.5$, QTG stars follow a slower radius growth, allowing ${\cal C} $ to exceed the black hole threshold as mass increases. This behavior is also observed for other values of the coupling constant $\lambda$ and for different EOSs~\cite{LD19666companion}.

This behavior can  be elucidated via a simple scaling argument.  For a uniformly dense star, the radius scales as $R \propto M^{1/3}$,  yielding a compactness  ${\cal C} \propto M^{2/3}$.  Thus, in the absence of an upper mass bound, as enabled by QTG, compactness can naturally exceed the black hole limit at sufficiently high mass. The precise mass at which this threshold is crossed depends sensitively on the EOS. Moderately soft EOSs like SLy typically produce super-compact stars only at high masses, whereas stiffer EOSs like DSQS~\cite{Gondek-Rosinska:2008zmv} allow the transition at comparatively lower masses.
 
Importantly, the uncertainty in the true EOS of neutron stars implies that a single EOS may support both conventional and ultra-compact stars within QTG, all consistent with current astrophysical constraints. For example, the $M-R$ relation in Fig.~\ref{fig:mrcrhoic} shows that for $\lambda = 100 \lambda_\star$ and the SQSB56 EOS~\cite{Gondek-Rosinska:2008zmv}, ultra-compact stars exceeding the black hole compactness arise at high central densities, while at lower densities, typical $1.4 M_\odot$ neutron stars remain fully compatible with stringent multi-messenger bounds~\cite{Capano:2019eae}. A more systematic exploration of the dependence of the stellar compactness on both the gravitational coupling and the EOS can be found in Ref.~\cite{LD19666companion}, which provides further evidence that the emergence of ultra-compact configurations is a robust feature of QTG, while the EOS mainly affects the quantitative location of the required central density.

Turning to the $I-{\cal C}$ relation in the right panel of Fig.~\ref{fig:mrcrhoic},  we observe another significant departure from GR. In GR, the moment of inertia increases with compactness only up to a certain point, beyond which it declines, a consequence of the mass-radius relation saturating near the black hole limit. In contrast, QTG stars display a monotonic and increasingly steep rise in $I$ with growing ${\cal C}$. This arises because, at high compactness, both mass and radius increase, leading to $I \propto M R^2$ becoming increasingly large. As a result, the angular velocity  $\Omega = J/I$ at fixed angular momentum $J$  is naturally suppressed for QTG stars, providing intrinsic dynamical stability against rotational instabilities. This structural behavior further reinforces the validity of the slow-rotation approximation adopted in our model construction. Notably, the enhancement of moment of inertia and compactness occurs without any fine-tuning of the EOS or additional fields, highlighting the intrinsic capacity of QTG to support super-compact, horizonless stellar configurations.

\textit{Stability} 
While both the exterior spacetime and global structure of stellar models in QTG support the existence of configurations exceeding the compactness of black holes, a fundamental question remains: are such ultra-compact stars stable under perturbations? As a first step toward addressing this, we analyze radial adiabatic perturbations of the static equilibrium configuration shown in Fig.~\ref{fig:solution}. 

Following the standard formalism~\cite{Chandrasekhar:1964zza,Misner:1973prb,Kokkotas:2000up,Li:2025gna}, we consider radial  perturbations of fluid elements inside the star, modeled by a  displacement  of the form $\delta r(r,t) = \xi(r) e^{i \omega t}$, where $\omega$ is the mode frequency. For the stellar model  shown in Fig.~\ref{fig:solution}, we find that the squared frequency of the fundamental mode is $\omega^2 = -4.2 \times 10^{-4} \omega_\star^2$ in GR and $\omega^2 = 1.3 \times 10^{-3} \omega_\star^2$ in QTG, where $\omega_\star$ sets the physical frequency scale~\cite{footnote1}. The negative value in GR indicates an exponentially growing  unstable mode, while the positive value in QTG corresponds to a stable, oscillatory mode. This demonstrates that the same stellar model, unstable and sub-black-hole compact in GR  becomes both more compact than a black hole and radially stable in QTG. A detailed analysis of radial stability is presented in Ref.~\cite{LD19666companion}.

While radial stability is a necessary condition, non-radial perturbations must also be addressed. Interestingly, the large moment of inertia associated with ultra-compact stars in QTG naturally leads to slower rotation at fixed angular momentum. Although previously identified as a structural feature, this also has important dynamical consequences: reduced rotation can substantially suppress the growth of non-radial instabilities commonly encountered in GR~\cite{Cardoso:2017cqb,Cardoso:2019rvt,Friedman:1978wla,Cardoso:2007az,Brito:2015oca,LISA:2022kgy}. Moreover, our stellar models are constructed without fine-tuning the matter content. This implies that stabilization mechanisms known from GR, such as those involving specific material properties or energy dissipation channels~\cite{Maggio:2017ivp,Maggio:2018ivz}, could remain effective within the QTG framework. Taken together with the confirmed radial stability, these considerations strongly suggest that ultra-compact stars in QTG are likely to be stable under a wide range of physically relevant perturbations.

\textit{Observational signature} 
Having established the viability and stability of ultra-compact stars within the QTG framework, we now explore  their potential astrophysical implications. Unlike conventional neutron stars, these ultra-compact stars can exhibit a much broader range of masses, making them plausible candidates for compact objects observed in astrophysical events~\cite{LIGOScientific:2020zkf,LIGOScientific:2024elc}, particularly those residing in the mass gap between the heaviest confirmed neutron stars and the lightest observed black holes. Moreover, since these stars are more compact than black holes yet free of singularities and event horizons, they may also serve as alternatives to certain astrophysical sources traditionally attributed to stellar-mass black holes, especially in light of the fact that event horizons have yet to be directly observed~\cite{Cardoso:2016rao,Cardoso:2016oxy,Cunha:2018gql,Herdeiro:2021lwl,Cardoso:2017cqb,Cardoso:2019rvt,Cunha:2018acu} and the existence of singularities is  generally considered an undesirable feature in any theory.

Given that the stars are more compact than black holes and possess photon spheres, despite lacking event horizons, their observational signals might closely resemble those from black holes or from other horizonless compact objects with the same mass but lower compactness (${\cal C} \le 0.5$). Therefore, it is crucial to identify distinctive signatures that can break this degeneracy through compactness-related effects. Such signatures would offer a unique probe of strong-field deviations from GR, particularly because these ultra-compact configurations can only arise from gravitational modifications. Within GR, pioneering studies by Cardoso et al.~\cite{Cardoso:2016rao,Cardoso:2016oxy} identified gravitational wave signals during the late-time ringdown stage of a binary coalescence as a promising means of determining whether extremely compact objects possess an event horizon.  For a regular object without an event horizon but compact enough to possess a clean photon sphere, gravitational waves generated during the early-time ringdown stage are repeatedly reflected between the object's surface (or interior) and the photon sphere. Each interaction at the photon sphere results in a portion of the wave escaping, producing a series of echoes with progressively decreasing amplitude during the late-time ringdown stage. In contrast, for a black hole, if the region just outside the event horizon is surrounded by an effective ``reflecting wall''\cite{Cardoso:2017cqb,Cardoso:2019rvt,Giddings:2017jts,Barausse:2014tra,Price:2017cjr,Mark:2017dnq}, potentially due to quantum effects or classical matter near the horizon, similar gravitational wave echoes could also be produced. A key distinction lies in the time delay between consecutive echoes. This delay depends on the distance from the surface of a regular object (or, in the case of a black hole with horizon-scale corrections, the effective reflecting wall) to its photon sphere. Since the photon sphere is typically located at $3M$, objects with higher compactness have a greater distance between their surface and the photon sphere, resulting in a longer time delay.

In QTG, although the exterior spacetime of the stellar model deviates from the GR prediction, the gravitational wave echo time delay $\Delta t$ and the corresponding echo frequency $2\pi/\Delta t$ can still be estimated using the standard formula $\Delta t \sim 2 \int_{R}^{R_{\textup{ps}}} (h f)^{-1/2} dr$~\cite{Cardoso:2016rao, Cardoso:2016oxy},  where $R$ is the stellar radius, $R_{\textup{ps}}$ the photon sphere location, and $h$, $f$ the metric functions. As shown in Table~\ref{echoes}, stellar models of the same mass but increasing compactness yield progressively lower echo frequencies, consistent with the behavior found in GR, where more compact configurations lead to longer travel times between the surface and the photon sphere.

\begin{table*}[t]
\centering
\caption{Gravitational wave echo frequencies for stars of the same mass $M = 6.84 M_{\odot}$ with a given EOS of SLy at different compactness levels.}\label{echoes}
\begin{tabular}{@{}lllll@{}}
\toprule
Radius ($r_\star$) & Compactness  & Echo frequency (Hz) & Central density ($\rho_\star$) & Coupling constant ($\lambda_\star$)  \\
\midrule
12.389    & 0.552   & 19497  & $2.534 \times 10^{-3}$ & 1000  \\
12.103    & 0.565   & 17468  & $3.125 \times 10^{-3}$ & 800  \\
11.934    & 0.573   & 16349  & $3.804 \times 10^{-3}$ & 650  \\
11.816    & 0.579   & 15500  & $4.857 \times 10^{-3}$ & 500  \\
\bottomrule
\end{tabular}
\end{table*}

Recent observational searches for echo signals have reported some positive results~\cite{Abedi:2016hgu,Conklin:2017lwb,Abedi:2021tti}. With the continued improvement in the sensitivity of existing detectors and the advent of next-generation interferometers~\cite{Punturo:2010zz,LIGOwhite,LISA:2017pwj,Reitze:2019iox}, conclusive evidence for these echoes may soon be possible.
Such a discovery would provide suggestive evidence for the existence of stars more compact than black holes and could potentially hint at modifications to gravity beyond GR.

\textit{Conclusions} 
Now we have shown that stars more compact than black holes, which are impossible within GR, can exist as regular, horizonless configurations within QTG. These stars are stable against radial perturbations and can produce distinct gravitational wave echo signatures, distinguishing them from less compact objects. Several directions for further investigation are warranted.

Firstly, it is natural to explore whether such ultra-compact stars can arise in other alternative gravity theories. The model~(\ref{qtg}) considered here represents a minimal case within a broader class of QTG theories. More generalized versions of QTG~\cite{Li:2017ncu,Hennigar:2017ego,Chen:2022fdi} may support a wider range of equilibrium solutions, including stars with even higher masses or more extreme compactness. Moreover, given the well-established equivalence between higher-curvature gravity theories and scalar-tensor or matter-coupled theories~\cite{Jakubiec:1988ef}, similar configurations might also emerge  in alternative formulations where gravity couples to dynamical fields. Notably, these theories may not reduce to GR directly in the weak-field limit as QTG does, but instead may require screening mechanisms --- such as the Vainshtein, chameleon, or symmetron effects --- to ensure consistency with observational data  in weak-field environments~\cite{Koyama:2015vza}.

Secondly, it is essential to investigate the behavior of these stars in more complex dynamical settings. Two key areas for further study include their evolution during realistic astrophysical processes, such as gravitational collapse or binary mergers, and their response to non-radial perturbations and other external disturbances~\cite{Keir:2014oka,Cardoso:2014sna,Cunha:2022gde}.  Addressing these issues requires a theory-sensitive approach, as perturbation dynamics are closely tied to the underlying gravitational framework. In extensions of GR such as QTG, the perturbation equations can differ significantly  from those of GR, and instabilities identified within GR may no longer apply. Conversely, QTG may introduce new stabilizing effects, as demonstrated by the radial perturbation stability shown earlier. Due to the higher-derivative nature of QTG, the spacetime evolution becomes a more intricate initial value problem. As a result, the dynamical outcomes may depend sensitively on the initial conditions and nonlinear interactions, rendering numerical simulations indispensable.  Importantly, even if instabilities develop and lead to disintegration, collapse into a black hole, or other forms of evolution, such scenarios do not imply any physical inconsistency within QTG. Unlike in GR, where the exterior spacetime remains fixed and unaffected by the star’s internal dynamics, the exterior in QTG evolves dynamically along with the stellar interior.

Lastly, several aspects of observational signatures warrant further exploration. On one hand, a more detailed study of gravitational wave echo characteristics would be invaluable. This could be achieved by extending the current work to consider various stellar properties and generalizing the QTG model to cover a broader frequency range. In addition, a comprehensive analysis of the dynamical response of these stars under perturbations is needed in order to extract detailed information about the echo waveform and frequency.
On the other hand, other measurements, such as tidal deformability~\cite{Cardoso:2017cfl}, tidal heating~\cite{Maselli:2017cmm}, and the multipolar structure~\cite{Krishnendu:2017shb} of  objects during the inspiral phase of a binary coalescence could provide valuable information for distinguishing the nature of these highly compact objects. While the absence of an event horizon in these stars  results in no tidal heating, these horizonless ultra-compact objects  allow for tidal deformability and possess distinct multipolar structures~\cite{Cardoso:2017cqb,Cardoso:2019rvt,Berti:2015itd,Barack:2018yly,LISA:2022kgy}. These observables are expected to vary significantly with compactness, and when combined with echo signals, could enable a more complete characterization of the structure of compact objects. 
As the sensitivity of both gravitational-wave and electromagnetic-wave detectors improves,  future observations may provide critical insights into the nature of compact objects and the need to go beyond Einstein's theory of gravity.

\appendix*  \label{appendix}

\textit{Appendix A: Schwarzschild geometry in QTG}  \label{appendixA}
By varying the Lagrangian~(\ref{qtg}) with respect to the metric $g^{\mu\nu}$, we obtain the vacuum gravitational field equations, which are given by
\begin{equation}
E_{\mu\nu}  \equiv {\cal P}_{\mu\alpha\beta\gamma}{{\cal R}_\nu}^{\alpha\beta\gamma}-\frac{1}{2}g_{\mu\nu}L-2\nabla^\alpha\nabla^\beta {\cal P}_{\mu\alpha\beta\nu} =0 \,, \label{eom}
\end{equation}
where
${\cal P}_{\mu\alpha\beta\gamma} \equiv {\partial L}/{\partial {\cal R}^{\mu\alpha\beta\gamma}}$, ${\cal R}^{\mu\alpha\beta\gamma}$ is the Riemann tensor, and $\nabla^\alpha$ represents the covariant derivative. The spacetime of a slowly rotating black hole can also be described by Eq.~(\ref{ansatz}).
By substituting Eq.~(\ref{ansatz}) into Eq.~(\ref{eom}) and performing a Taylor expansion in the slow-rotation parameter $\epsilon$, we obtain two independent nonlinear higher-derivative ordinary differential equations (ODEs) for $h$ and $f$ at ${\cal O}(\epsilon^0)$:
\begin{eqnarray}
F_1(r, h^{\prime\prime\prime}, h^{\prime\prime}, h^{\prime}, h, f^{\prime\prime}, f^{\prime}, f) &=& 0 \,, \label{eom1} \\
F_2(r, f^{\prime\prime\prime}, h^{\prime\prime}, h^{\prime}, h, f^{\prime\prime}, f^{\prime}, f) &=& 0 \,, \label{eom2} 
\end{eqnarray}
and one linear fourth-order ODE for $w$ at ${\cal O}(\epsilon^1)$:
\begin{equation}
F_3(r, w^{\prime\prime\prime\prime}, w^{\prime\prime\prime}, w^{\prime\prime}, w^{\prime}, h^{\prime\prime}, h^{\prime}, h, f^{\prime\prime}, f^{\prime}, f) = 0 \,, \label{eom3} 
\end{equation}
where a prime denotes the derivative with respect to $r$, and the $F_i$ encapsulate combinations of the metric functions and their derivatives, simplifying the notation.

Let $h = f = 1 - 2 M / r$, and substitute them into Eqs.~(\ref{eom1})-(\ref{eom2}). It can be readily verified that the Schwarzschild metric satisfies the vacuum gravitational field equation at ${\cal O}(\epsilon^0)$ within the framework of QTG. Furthermore, substituting the same expressions into Eq.~(\ref{eom3}) reduces it to a second-order ODE for $w$, whose solution is given by $w = \Omega - 2 J/r^3$, identical to the result in GR. This confirms the existence of slowly rotating Schwarzschild black holes and establishes that the Schwarzschild geometry corresponds to a maximally symmetric vacuum in the weak-field regime.

For completeness, we briefly review the linearized theory of QTG to demonstrate that it reduces to GR in the weak-field regime, as previously established~\cite{Li:2017ncu, Hennigar:2017ego}. The exterior vacuum spacetime of a slowly rotating compact object, at sufficiently large distances from its boundary, can be approximated as a linear perturbation to flat spacetime, expressed as $g_{\mu\nu} = \bar{g}_{\mu\nu} + \tilde{g}_{\mu\nu}$, where ``bar'' and ``tilde'' denote quantities associated with the background and perturbations, respectively.  For the maximally symmetric backgrounds, the Riemann tensor takes the form $\bar{{\cal R}}_{\mu\nu\rho\sigma} = K ( \bar{g}_{\mu\rho}\bar{g}_{\nu\sigma} - \bar{g}_{\mu\sigma}\bar{g}_{\nu\rho})$, with $K$ a constant. The linearized vacuum gravitational field equation in QTG is given by~\cite{Li:2017ncu, Hennigar:2017ego}
\begin{equation}
 \tilde{E}_{\mu\nu} \equiv \delta \tilde{{\cal R}}_{\mu\nu} -\frac12 \bar{g}_{\mu\nu}  \delta \tilde{{\cal R}} =0 \,, \label{weak}
\end{equation} 
with $\delta \tilde{{\cal R}}_{\mu\nu} =  \tilde{{\cal R}}_{\mu\nu} - K \tilde{g}_{\mu\nu}$,  and  $ \delta \tilde{{\cal R}} = \bar{g}^{\mu\nu} \delta \tilde{{\cal R}}_{\mu\nu} $\,. These equations align entirely with those of GR in weak-field regimes, as if QTG behaves as a purely topological theory in this limit. Note that, although the full vacuum gravitational field equations~(\ref{eom}) are higher-derivative in nature, they reduce to a second-order form in the weak-field regime. Consequently, the resulting solutions contain only a single integration constant --- interpreted as the mass --- mirroring the GR case. This confirms that, in QTG, the linearized weak-field spectrum comprises only the standard massless graviton, with no scalar or ghost-like massive spin-2 modes~\cite{Stelle:1976gc}.

Next, to investigate whether QTG admits black hole solutions that deviate from the Schwarzschild spacetime at ${\cal O}(\epsilon^0)$, we numerically solve Eqs.~(\ref{eom1})-(\ref{eom2}). The first step involves determining the boundary conditions at the event horizon and at infinity. For a slowly rotating black hole described by metric~(\ref{ansatz}) with a well-defined event horizon at radius $r_0$, where the metric functions $h$ and $f$ vanish simultaneously, the near-horizon behavior of $h$ and $f$ can be expressed as a Taylor expansion~\cite{Lu:2015cqa}
\be
\lim_{r\rightarrow r_0}h(r) = \sum_{r\rightarrow r_0}h_n (r- r_0)^n \,, \lim_{r\rightarrow r_0}f(r) = \sum_{r\rightarrow r_0}f_n (r- r_0)^n \,, \label{nearhorizon}
\ee
where $h_1$ is a trivial parameter, reflecting the freedom to rescale the temporal coordinate. Substituting Eq.~(\ref{nearhorizon}) into Eqs.~(\ref{eom1})-(\ref{eom2})  yields two distinct branches of boundary conditions at the event horizon. The first few coefficients $h_i$ and $f_i$ for each case are:
\bea
\hspace{-1cm}&&h_2 = -\frac{1}{r_0} \,,\quad f_1 = \frac{1}{r_0} \,,\quad f_2 = -\frac{1}{r_0^2}  \,,  \label{case1} \\
\hspace{-1cm}&&h_2 = -\frac{1}{r_0} +\frac{r_0^3}{24\lambda} \,,\quad  f_1 = \frac{1}{r_0} \,,\quad f_2 = -\frac{1}{r_0^2} -\frac{r_0^2}{8\lambda}  \,. \label{case2}
\eea
At infinity, the large-$r$ behavior satisfying the asymptotic flatness condition must correspond to the Schwarzschild metric:
\begin{equation}
\lim_{r\rightarrow\infty}h(r) = \lim_{r\rightarrow\infty}f(r) = 1 - \frac{2 M}{r} \,, \label{infinity}
\end{equation}
where $M$ represents the mass. For both cases,  the boundary conditions at the event horizon depend solely on the parameter $r_0$, with no additional free parameters in the coefficients $h_i$ and $f_i$. [Taylor expansions are computed up to ${\cal O}((r - r_0)^{19})$.] Consequently, the solutions for case I (\ref{case1}) and case II (\ref{case2}) represent two bifurcating branches, with no integration constants interpolating between them.

\begin{figure}[ht]
\centering
\includegraphics[width=0.75\linewidth]{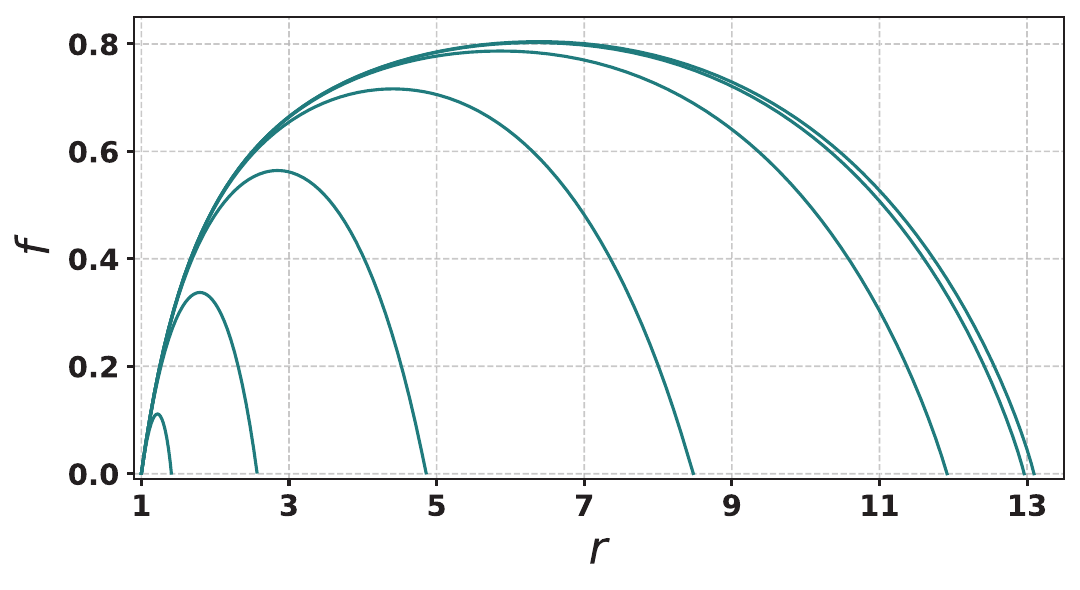}
\caption{\label{fig:mrsly}The solutions of metric function $f$ for Eqs.~(\ref{eom1})-(\ref{eom2}) with the boundary condition~(\ref{case2}) at event horizon as initial conditions.  The radius of horizon is $r_0 = 1$ and the values of $\lambda$ of the curves from left to right are chosen as $(0.1, 1, 10, 10^2, 10^3, 10^4, 10^5)$.}\label{fplots}
\end{figure}

The boundary conditions (\ref{case1}) and (\ref{infinity}) are identical to those of the Schwarzschild black hole. If another well-defined black hole exists, deviating from the Schwarzschild geometry, its near-horizon and large-$r$ boundary conditions would correspond to (\ref{case2}) and (\ref{infinity}), respectively. However, numerical integration of the vacuum field equation (\ref{eom1}) and (\ref{eom2}) using the boundary conditions (\ref{case2}) at the event horizon as initial conditions fails to converge to the boundary conditions (\ref{infinity}) at infinity. Instead, the function $f(r)$ invariably decays to zero rather than asymptotically approaching 1, regardless of any finite value of $\lambda>0$, as illustrated in Fig.~\ref{fplots}. This indicates that in QTG, only one set of boundary conditions, (\ref{case1}) and (\ref{infinity}), leads to a well-defined black hole solution, which corresponds to the Schwarzschild black hole at ${\cal O}(\epsilon^0)$. Moreover, substituting the Schwarzschild metric functions $h$ and $f$ into the ${\cal O}(\epsilon^1)$ equation (\ref{eom3}), the equation reduces to a second-order linear ODE for $w$, identical to that in GR. This confirms that QTG admits the same slowly rotating Schwarzschild black hole solution as GR.

Consequently, from a non-perturbative perspective, we use numerical methods to confirm that the slowly rotating Schwarzschild black hole is the unique slowly rotating black hole solution within the framework of QTG. It is noteworthy that the property whereby ``vacuum solutions are not unique, while the black hole solution is'' is not exclusive to QTG; it is also a feature shared by many modified gravity theories~\cite{Lu:2015cqa, Liu:2020yqa}, including the well-known Starobinsky model~\cite{Starobinsky:1980te} and its equivalent scalar-tensor gravity.

\begin{figure}[ht]
\centering
\includegraphics[width=0.45\textwidth]{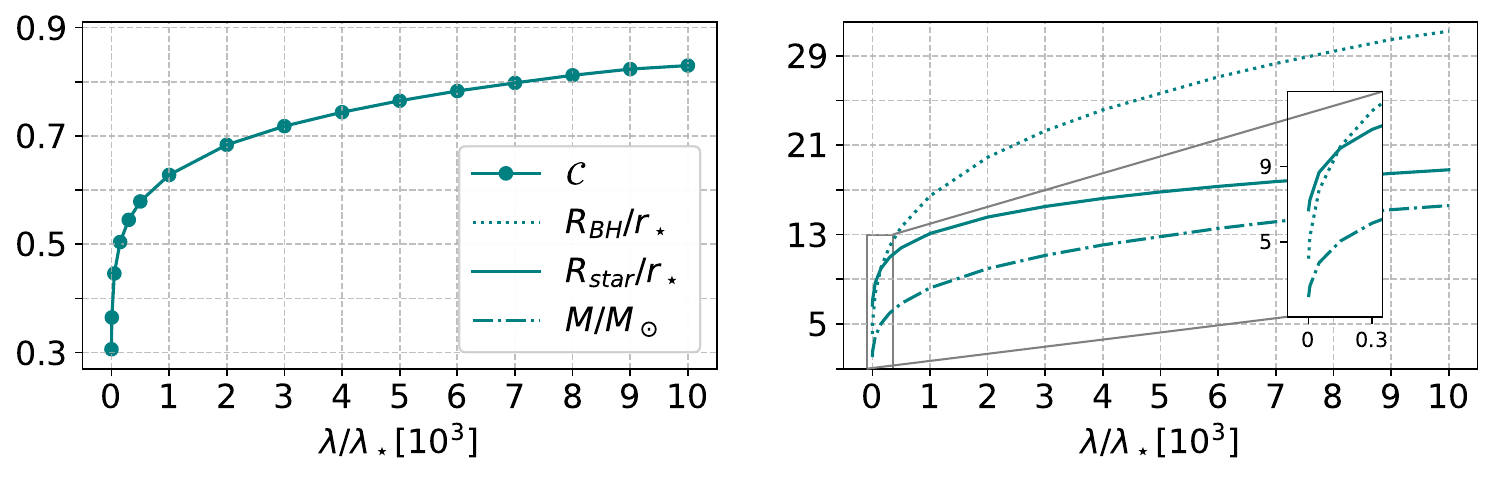}
\caption{The influence of QTG on the compactness $\mathcal{C}$, mass $M$, and radius $R_{\textup{star}}$ of neutron stars.  The left panel shows the ${\cal C}-\lambda$ relation, while the right panel presents the corresponding $M-\lambda$ (dash-dotted line) and $R_{\textup{star}}-\lambda$ (solid line) relations. The stellar model is constructed using the SLy EOS with a central density $\rho_0 = 4.86 \times 10^{-3} \rho_\star$. For comparison, the radius $R_{\textup{BH}}$ (dotted line) of black holes with the same mass $M$ is also shown.}\label{fig:clambda}
\end{figure}

\textit{Appendix B: Enhancement of compactness induced by QTG}
As discussed in the main text, the inclusion of QTG leads to an increase in both the mass and radius of a given stellar model. However, since the mass grows more rapidly than the radius, the compactness inevitably exceeds that of a black hole. To verify this trend, Fig.~\ref{fig:clambda} shows the compactness ${\cal C}$, mass $M$, and radius $R_{\textup{star}}$ of the stellar model depicted in Fig.~\ref{fig:solution} as functions of the coupling constant $\lambda$. For comparison, we also include the radius $R_{\textup{BH}}$ corresponding to QTG black holes of the same mass.

\textit{Acknowledgments}
We are grateful to Yifan Chen, Xing-Hui Feng, Chengjie Fu, Hyat Huang, Hong-Bo Li, Robert B. Mann, Samuele Silveravalle, Hao Wei, Puxun Wu, Shuang-Qing Wu, Yu-Peng Zhang and  Zhen Zhong for useful discussions. S.L. also thanks Lijing for the warm hospitality during the early stage of this work at PKU-KIAA.
S.L. and H.Y. were supported in part by the National Natural Science Foundation of China (12105098, 12481540179, 12075084, 11690034, 11947216, 12005059), Natural Science Foundation of Hunan Province (2022JJ40264), and the innovative research group of Hunan Province (2024JJ1006). H.L. was supported in part by the National Natural Science Foundation of China (11935009, 11875200). Y.G., R.X. and L.S. were supported in part by the National SKA Program of China (2020SKA0120300), and the Max Planck Partner Group Program funded by the Max Planck Society.


\begin{thebibliography}{99}
\bibitem{footnote1}
The characteristic dimensions of relevant physical quantities, normalized by the solar mass $M_{\odot}$ and expressed in centimeter-gram-second (CGS) units, are given by: $r_\star = G M_{\odot}/c^2 = 1.47 \times 10^{5} \textup{cm}$, $\rho_\star = c^6/(G^3 M_{\odot}^2 ) = 6.18\times 10^{17} \textup{g} \cdot \textup{cm}^{-3} $, $p_\star = c^8/(G^3 M_{\odot}^2 ) =5.55\times 10^{38}  \textup{g} \cdot \textup{cm}^{-1} \cdot \textup{s}^{-2}$,  $I_\star = G^2 M_{\odot}^3/c^4 = 4.34\times 10^{43}  \textup{g} \cdot \textup{cm}^{2} $, $  \lambda_\star = G^4 M_{\odot}^4/c^8 = 4.76 \times 10^{20} \textup{cm}^4 $, $\omega_\star = c^3/(G M_{\odot}) = 2.03 \times 10^{5} \textup{s}^{-1}$.
\end{thebibliography}

\end{document}